\documentclass[journal=ancac3,manuscript=article]{achemso}
\usepackage[version=3]{mhchem} 
\usepackage{xcolor}


\author{Chetana Badala Viswanatha}
\affiliation[Unknown University]
{Department of Physics, University of Kaiserslautern-Landau, Erwin-Schrödinger-Straße 46, 67663 Kaiserslautern, Germany}
\alsoaffiliation[Second University]
{Research Center OPTIMAS, Erwin-Schrödinger-Straße 46, 67663 Kaiserslautern, Germany}
\alsoaffiliation[]{Department of Physics, Prayoga Institute of Education Research, Bengaluru 560116, India}
\email{chetana.bv@prayoga.org.in}
\author{Johannes Stöckl}
\affiliation[Unknown University]
{Department of Physics, University of Kaiserslautern-Landau, Erwin-Schrödinger-Straße 46, 67663 Kaiserslautern, Germany}
\alsoaffiliation[Second University]
{Research Center OPTIMAS, Erwin-Schrödinger-Straße 46, 67663 Kaiserslautern, Germany}
\author{Benito Arnoldi}
\affiliation[Unknown University]
{Department of Physics, University of Kaiserslautern-Landau, Erwin-Schrödinger-Straße 46, 67663 Kaiserslautern, Germany}
\alsoaffiliation[Second University]
{Research Center OPTIMAS, Erwin-Schrödinger-Straße 46, 67663 Kaiserslautern, Germany}
\author{Johannes Knippertz}
\affiliation[Unknown University]
{Department of Physics, University of Kaiserslautern-Landau, Erwin-Schrödinger-Straße 46, 67663 Kaiserslautern, Germany}
\alsoaffiliation[Second University]
{Research Center OPTIMAS, Erwin-Schrödinger-Straße 46, 67663 Kaiserslautern, Germany}
\author{Florian Haag}
\affiliation[Unknown University]
{Department of Physics, University of Kaiserslautern-Landau, Erwin-Schrödinger-Straße 46, 67663 Kaiserslautern, Germany}
\alsoaffiliation[Second University]
{Research Center OPTIMAS, Erwin-Schrödinger-Straße 46, 67663 Kaiserslautern, Germany}
\author{Hauke Paulsen}
\affiliation[Unknown University]
{Institute of Physics, University of Lübeck, Ratzeburger Allee 160, 23562 Lübeck, Germany}
\author{Juliusz A. Wolny}
\affiliation[Unknown University]
{Department of Physics, University of Kaiserslautern-Landau, Erwin-Schrödinger-Straße 46, 67663 Kaiserslautern, Germany}
\author{Martin Aeschlimann}
\affiliation[Unknown University]
{Department of Physics, University of Kaiserslautern-Landau, Erwin-Schrödinger-Straße 46, 67663 Kaiserslautern, Germany}
\alsoaffiliation[Second University]
{Research Center OPTIMAS, Erwin-Schrödinger-Straße 46, 67663 Kaiserslautern, Germany}
\author{Volker Schünemann}
\affiliation[Unknown University]
{Department of Physics, University of Kaiserslautern-Landau, Erwin-Schrödinger-Straße 46, 67663 Kaiserslautern, Germany}
\author{Benjamin Stadtmüller}
\affiliation[Unknown University]
{Department of Physics, University of Kaiserslautern-Landau, Erwin-Schrödinger-Straße 46, 67663 Kaiserslautern, Germany}
\alsoaffiliation[Second University]
{Research Center OPTIMAS, Erwin-Schrödinger-Straße 46, 67663 Kaiserslautern, Germany}
\alsoaffiliation{Experimentalphysik II, Institute of Physics, Augsburg University, 86159 Augsburg, Germany}
\email{benjamin.stadtmueller@uni-a.de}

  \title[An \textsf{achemso} demo]
  {Impact of Cooperativity on the Spatial and Temporal Evolution of the Light-induced Spin-State Switching of the Fe(phen)$_2$(SCN)$_2$ Spin-Crossover Complex}


\begin{document}

\begin{tocentry}

  \includegraphics[width=\textwidth]{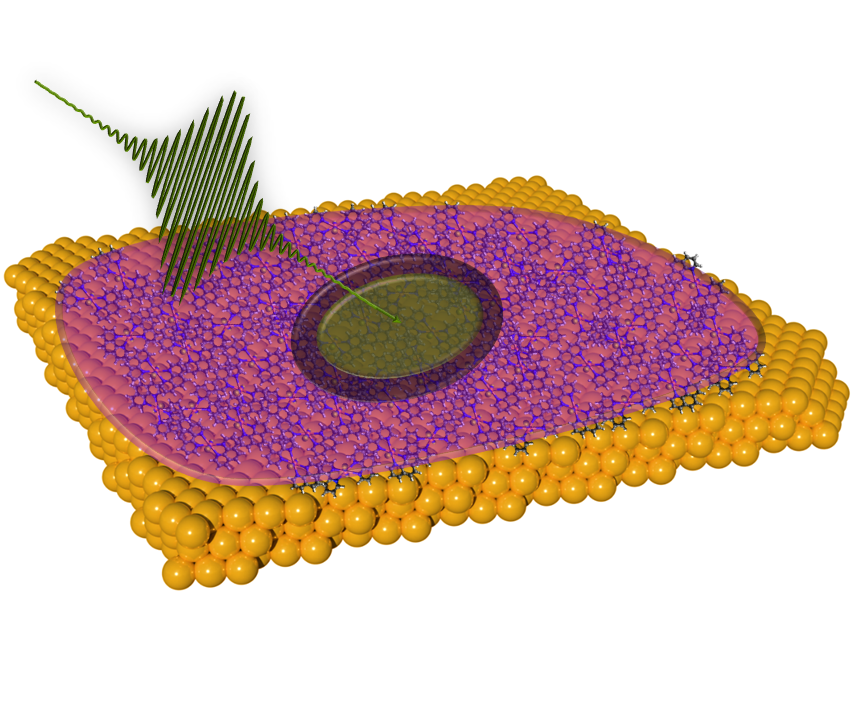}

\end{tocentry}

\begin{abstract}
Spin crossover (SCO) complexes are highly flexible bistable molecular switches with two distinct spin states that can be switched into each other by external stimuli such as temperature, pressure, or light. In the condensed phase, this spin switching phenomenon is determined not only by the chemical and structural properties of the SCO compound but also by intermolecular interactions within the SCO material and their coupling to surfaces. These interactions lead to cooperative effects in the spin switching behavior that are, for instance, reflected in the thermal hysteresis of the spin switching or determine the speed of the thermal and optical spin transition within a thin film of SCO complexes. 

In this study, we shed new light on the cooperativity-driven spatial and temporal dynamics of the collective light-induced spin switching of thin SCO films on a gold single crystal. We quantify the spatial and temporal dynamics of the transition between the low-spin and high-spin \textit{light-induced excited spin state trapping} (LIESST) states of the Fe(phen)$_2$(SCN)$_2$ compound. Using real-time photoemission electron microscopy with millisecond time resolution, we uncover spectroscopic signatures of the light-induced spin state switching to the LIESST state even $200\, \mu$m outside the spot of direct optical excitation. This observation is identical for both continuous wave and femtosecond laser light excitations in the visible spectral range. The timescale of the spin-switching dynamics is indicative of a photothermal process in which molecules far from the laser spot relax back to the low-spin ground state faster than those within the directly illuminated region. All of these observations highlight the crucial role of cooperative interactions in the spin state switching of the SCO thin film. 

\end{abstract}


\section*{Introduction}
The continuing miniaturization of device structures in quantum and information technology requires the development and understanding of novel materials with functionalities that can be exploited at the nanoscale. In this endeavor, spin crossover (SCO) complexes have emerged as highly flexible molecular spin switches with intriguing functionalities for applications \cite{Ruben+2017,Kamilya+2024}. SCO complexes consist of a transition metal center ion - in most cases an iron center - surrounded by organic ligands that can be reversibly switched between a $S=0$ low spin (LS) and a $S=2$ high spin (HS) state by external stimuli such as temperature, pressure, currents, electric fields, and (ultrashort) light pulses. This behavior can be observed in a manifold of different condensed matter systems, such as single molecules in break junction geometries, molecular monolayers and thin films on surfaces, and molecular crystals with qualitative and quantitative differences in the switching behavior for the different systems. In all these cases, the switching of the spin state is accompanied by changes in other properties of the material such as electronic structure, magnetic response, or molecular volume \cite{spintronics2011, spintronics2013, isomolthinfilm, spintronics5}, making SCO complexes a versatile molecular switch for applications in nanoscale quantum and information technology, sensing, and spintronics \cite{spintronics2014, spintronics2018, spintronics4, molecularswitch}. 

Despite recent progress in understanding the spin-state transition in selected classes of SCO molecules, a general picture of the SCO phenomenon in the condensed matter phase is still elusive. One reason is the complex interplay of intra- and intermolecular interactions that govern the SCO transition in bulk materials or thin films grown on surfaces. In particular, intermolecular interactions within SCO films can lead to cooperative effects\cite{cooperativeshort-range, cooperativebistability, cooperativetheory, cooperativescolatestreview} that dictate how individual molecular properties influence and are influenced by their neighboring molecules. These interactions are shaped by intrinsic molecular properties, the molecular structure, substituents, and packing density in thin films, resulting in spin-switching behavior that differs from that of isolated molecules. For example, cooperative interactions such as hydrogen bonding, $\pi$-$\pi$ stacking, and van der Waals forces can not only alter the SCO transition temperature and hysteresis \cite{cooperativeorganoborate, cooperativeimp}, but can also lead to thermally induced trapping of the high spin state at low temperatures \cite{Kiehl2022}.

In addition, cooperative effects are also crucial for the temporal dynamics of the switching process itself \cite{dynamicschiral, dynamicsimp}. In particular, such effects can alter the kinetics of the spin transition, including the rate and stability of the transition\cite{dynamicsmemorydev, kineticsdynamics}. For example, they can either speed up or slow down these transitions\cite{LIESSTcooperative}, thus affecting the response time and stability of externally stimulated spin functionalities in a potential SCO-based device structure.

In this work, we investigate the role of cooperativity in the light-induced spin-state switching of thin films of a Fe-based SCO complex on a gold surface. Three distinct steps govern this process: First, the rapid photo-switching leads to a local change of the molecular spin state at individual sites within a few ps \cite{laserheat2, laserheat1}. This local spin transition is then imprinted onto the neighboring molecules by lattice expansion (ns-time scale) and finally to the whole molecular film by thermal HS-state occupation (ms and longer). These different steps lead to the SCO process's characteristic spatial and temporal dynamics, which must reflect characteristic signatures of the cooperative coupling within the SCO material. However, a complete investigation of the spatio-temporal dynamics of the entire switching process is extremely challenging due to the many different timescales involved in the transitions. 

Here, we focus on the spatio-temporal dynamics of the last step, i.e., the thermal SCO switching from the LS into the HS state in the whole SCO film after optical excitation with light. For the selected SCO complex  Fe(phen)$_2$(SCN)$_2$, optical excitation with continuous wave as well as ultrashort light pulses can trigger a transition from the LS state to an HS-like Light-Induced Excited Spin State Trapping (LIESST) state\cite{LIESST, LIESST2019}, which can persist for several seconds up to several hours after excitation before relaxing back to the LS state. In our study, we use photoemission spectroscopy (PES) to identify the spectroscopic signatures of the LS and HS states\cite{stmxas,stmxas2,visvuv,XrayPEEM, stm, uhvsco1, JPCLrecent, uvarpessco} for different film thicknesses. In addition, we employ vacuum ultraviolet photoemission electron microscopy (VUV-PEEM) to monitor the spatial and temporal dynamics of the light-induced spin-state switching with high spatial resolution down to $100\,$nm. 

For thin films of the prototypical SCO complex Fe(phen)$_2$(SCN)$_2$ on an Au(111) single crystal, we observe that the light-induced LS to LIESST state switching extends up to $200\,\mu$m away from the laser illuminated area, illustrating the crucial role of cooperative interactions for the SCO transition for this material. The timescale of these dynamics indicates a photothermal process in which molecules far from the laser spot relax back to the low-spin state faster than those within the laser spot, providing valuable insights into the cooperative phenomena influencing SCO transitions.

\section*{Results and Discussion}

\begin{figure}[h!]
\centering
  \includegraphics[height=10.5cm]{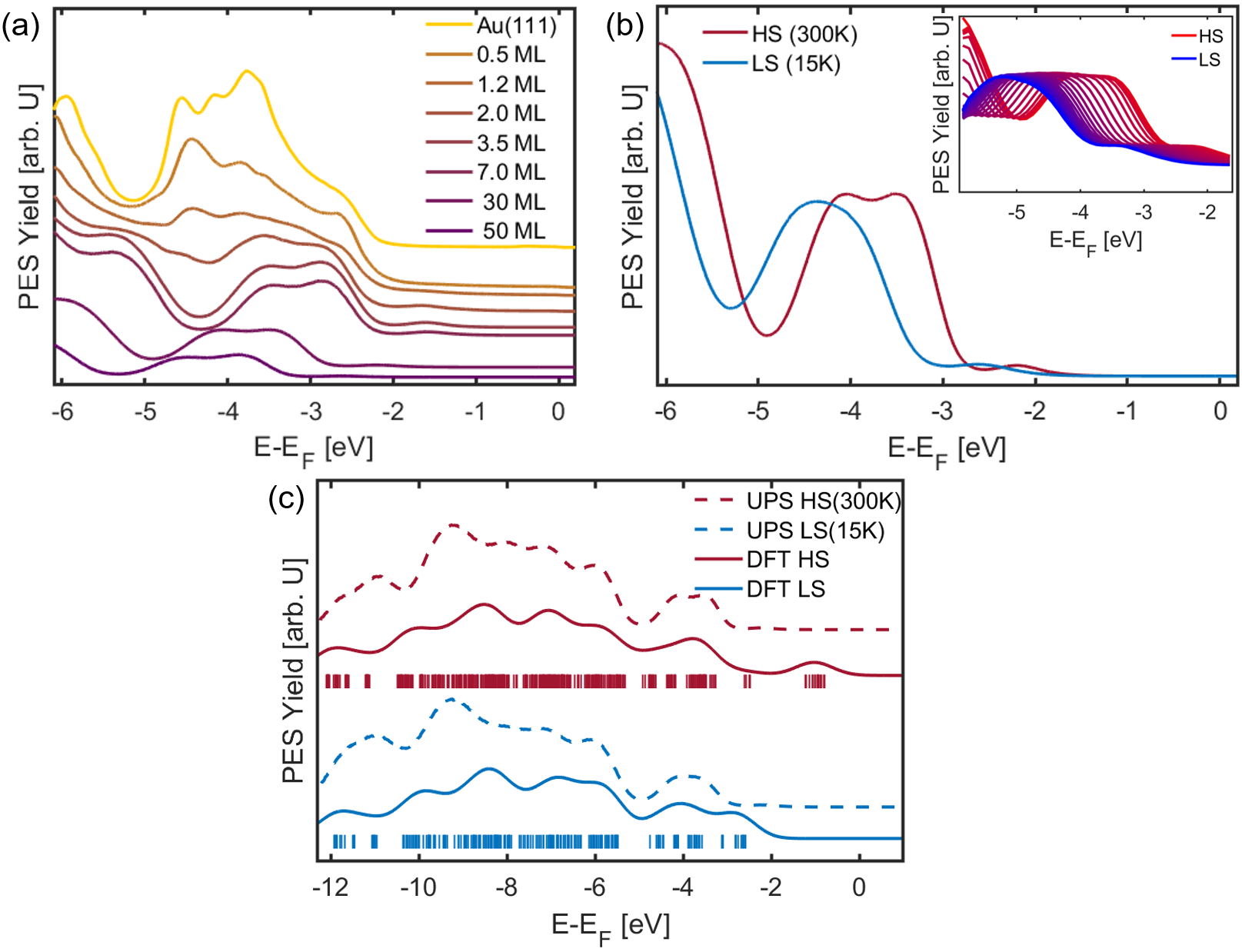}
  \caption{(a) Photoemission valence band spectra of the high spin state of thin Fe(phen)$_2$(SCN)$_2$ films with different film thicknesses grown on gold. The spectra were recorded in the normal emission geometry and at room temperature. The spectra have been normalized and vertically shifted for better visibility. A spectrum of the bare gold surface is included for reference. (b) Photoemission valence band spectra of the LS (blue solid line) and HS (red solid line) states of a $30\,$ML thick Fe(phen)$_2$(SCN)$_2$ film recorded at $15\,$K and $300\,$K, respectively. The inset shows a series of similar photoemission spectra recorded for a $40\,$ML film around its SCO transition temperature. The spectra are shown as a mixture of red and blue, where the increasing contribution of blue indicates a decrease in sample temperature. (c) Comparison of experimental and simulated HS and LS state photoemission spectra for a $40\,$ML thick Fe(phen)$_2$(SCN)$_2$ film. The simulated spectra were obtained from the eigenvalues of a DFT calculation of six Fe(phen)$_2$(SCN)$_2$ molecules \cite{paulsendft} artificially broadened by a Gaussian functions.}
  \label{figure1}
\end{figure}

We begin our discussion by analyzing the spectroscopic signatures of the HS and LS states in our spatially integrated photoemission spectra. We start with our room-temperature photoemission data, representing the HS state's spectroscopic signatures. 
Figure~\ref{figure1}(a) shows the photoemission spectrum of the Au(111) substrate as a yellow solid line, accompanied by spectra of Fe(phen)$_2$(SCN)$_2$ thin films with different film thicknesses. The increasing purple color of the spectra corresponds to increasing film thickness. 

The spectra for ultrathin films ($0.5$ - $2\,$ML) are dominated by the 5d-band signals of the Au substrate and show only marginal contributions from the molecular layer. Interestingly, the gold surface state disappears completely, even for very small coverages of $0.5\,$ML (not shown here), indicating a homogeneous distribution of SCO molecules on the Au surface. Furthermore, any changes in the spectral line shape of the Au d bands in the energy range between $-2.2\,$eV and $-6.2\,$eV are mainly attributed to photoelectrons scattered at the molecular layer \cite{auscattering1, auscattering2, JSpaper}. 

Clear and distinct molecular signals can only be observed for thicknesses larger than $2\,$ML. For $\Theta=3\,$ML, the highest occupied molecular orbital (HOMO) is located at E-E$_\mathrm{F}=-1.6\,$eV, the following molecular orbitals (HOMO-1/2/3) at $-2.9\,$eV, $-3.6\,$eV, and $-5.3\,$eV, respectively. These peaks continuously increase in intensity with increasing film thickness up to $\approx 5\,$ML while the contributions from the gold substrate decrease. Afterwards, the intensity of the molecular signals does not increase substantially due to the high surface sensitivity of the corresponding small inelastic mean free path of the detected photoelectrons \cite{Graber+2011}. 

Increasing the film thickness beyond $7\,$ML leads to a continuous shift of the entire spectrum by up to $1\,$eV towards larger binding energies (E$_B=|E-E_\mathrm{F}|$) for coverages up to $50\,$ML without a significant change in the line shape of the individual peaks. Accordingly, we attribute this shift to a final state effect in the photoemission processes caused by the reduced dielectric screening of the photo-hole for large film thicknesses \cite{screeningenergyshifts}. 

The spectroscopic signatures of the LS state can now be identified by directly comparing photoemission spectra at temperatures above and below the transition temperature. Fig.~\ref{figure1}(b) shows the corresponding spectra of a $30\,$ML Fe(phen)$_2$(SCN)$_2$ at room temperature, i.e, in the HS state, as well as at $\approx15\,$K, which is significantly lower than the SCO transition temperature of this complex \cite{origin1,origin2}. Reducing the sample temperature below the transition temperature leads to a rigid shift of the valence band features by $\approx 0.3\,$eV to larger binding energies. 
This shift almost completely recovers when the sample is reheated to room temperature as will be discussed in more detail later. In addition to the spectral shifts, we observe an increase in the linewidth of the HOMO-1 and HOMO-2 emission lines at $15\,$K, which persists even when the sample is returned to room temperature. This subtle permanent change suggests a small irreversible structural change within the film due to its exposure to VUV radiation.

To further elucidate the spectroscopic signature of the temperature-driven HS-LS transition, we have recorded photoemission spectra of the $40\,$ML Fe(phen)$_2$(SCN)$_2$ film between room temperature (RT) to $140\,$K. Fig.~\ref{figure1}(b) shows the corresponding data as inset. They were taken in real-time as the sample was cooled, and therefore for equal time steps (but not temperature differences) between subsequently recorded spectra. 
The RT spectrum of the HS state is shown in red, the low-temperature (LT) spectrum of the LS state is shown in blue, and the spectra of the intermediate temperatures are shown as a mixture of red and blue colors. We find a continuous shift of all spectral signals of the entire valence band spectra towards larger binding energies with decreasing temperature. On the one hand, this shows that our photoemission data reflect the continuous nature of the SCO phase transition of the Fe(phen)$_2$(SCN)$_2$ complex. On the other hand, the continuous shift of all valence states is inconsistent with a simple linear superposition of the HS and LS state spectra that might be expected due to the changing ratio of HS and LS molecules during the phase transition. Instead, we propose that the continuous shift is not governed by the properties of individual molecules within the SCO film, but rather by a collective phenomenon caused by changes in the dielectric screening during the SCO phase transition. We attribute this to the different molecular volumes of the SCO complexes in the LS and HS states, which is associated with the different polarizability of the switched molecules within the film \cite{xrd}. As the number of switched molecules within the film gradually increases, this can lead to a gradual change in the polarizability of the entire film, resulting in a continuous shift of the entire valence band state. Therefore, we use the binding energy position of the molecular signatures (HOMO, HOMO-1/2) or the corresponding change in the vacuum level as a measure of the SCO transition.  

This conclusion is indirectly supported by density functional theory (DFT) calculations, which do not consider collective phenomena within the molecular film. In Fig.~\ref{figure1}(c), we compare the experimental and simulated photoemission spectra of the HS and LS state. The simulated spectra were obtained by a DFT calculation of six Fe(phen)$_2$(SCN)$_2$ molecules with periodic boundary conditions to approximate the conditions of a molecular solid \cite{paulsendft}. The energy eigenvalues of the DFT calculations were artificially broadened by Gaussian functions and summed to the valence band spectra of Fig.~\ref{figure1}(c). The experimental data correspond to those of the $30\,$ML Fe(phen)$_2$(SCN)$_2$ film discussed earlier. The experimental spectrum of the LS state has been energetically shifted by $\approx 0.5\,$eV for better comparison with the calculated spectrum. 

While the simulated spectra can describe the general line shape of the HOMO-X features in the energy range between E-E$_\mathrm{F}=-12\,$ and $-3\,$eV, there are striking differences between the simulation and the experiment regarding the signatures of the SCO transition. The most significant difference is the additional spectral density at E-E$_\mathrm{F}=-1\,$eV in the simulated HS spectrum. This signature is missing in the simulated LS spectrum and our experimental data for both the HS and LS states. Such a spectral density has been observed for thin films of other SCO complexes by photoemission experiments \cite{visvuv,Rohlf+2021} and has been identified as the occupation of the (single particle) d-states of the Fe center ion in the HS state. Here, we can only speculate about the absence of similar signatures in the HS spectra for the Fe(phen)$_2$(SCN)$_2$ thin films, which could be caused, for example, by an extremely small cross-section of the corresponding states under our experimental conditions. The second significant difference between the experimental and theoretical spectra is the absence of the valence band shift between the HS and LS spectra in the theoretical data. Overall, this comparison strengthens our conclusion that the continuous shift of the valence band states during the gradual HS to LS transition is caused by a change in the dielectric environment that is not correctly described at the standard DFT level.

\begin{figure}[t]
 \centering
 \includegraphics[height=9cm]{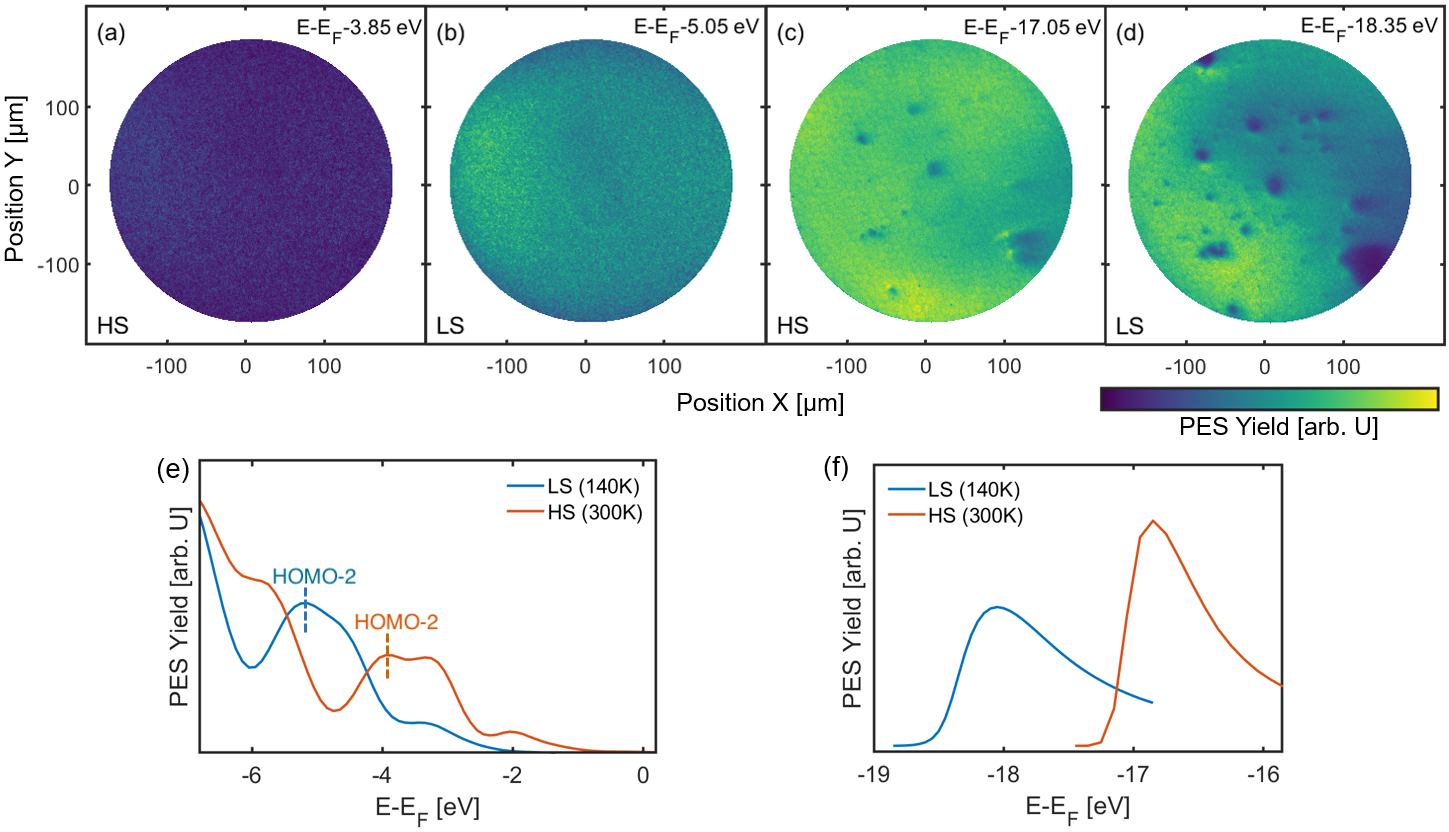}
 \caption{(a)-(d) Spatially resolved photoemission yield (PEEM image) of a $40\,$ML Fe(phen)$_2$(SCN)$_2$ film on gold in the HS state at room temperature (a)/(c) and in the LS state at $140\,$K (b)/(d), recorded for two characteristic spectral features. The images in (a)/(b) were obtained for the binding energy of the HOMO-2 state, and those in (c)/(d) for the low energy cutoff of the valence band spectrum. The energies of the features were adjusted for the HS and LS states to account for the energy shift of the valence band state during the temperature-induced spin transition. The corresponding photoemission spectra of the valence band and the cutoff region for both temperatures are shown in panels (e) and (f).}
 \label{figure2}
\end{figure}

Having identified the spectroscopic signatures of the HS and LS states, we now discuss the spatial and temporal dynamics of light-induced spin-state switching on long-time scales. For this purpose, we use a photoemission electron microscope (PEEM) equipped with an imaging energy filter to acquire energy-resolved images of the photoemission distribution of the sample surface with a sub-millisecond frame rate after optical excitation with continuous-wave or ultrashort light pulses. For these investigations we have chosen a $40\,$ML film of Fe(phen)$_2$(SCN)$_2$ on Au(111).

Figures.~\ref{figure2}(a)/(c) and (b)/(d) show the spatially resolved photoemission yield (or PEEM images) of the HS and LS states for the emission energies of the HOMO-2 state and the low energy cutoff of the sample. The HOMO-2 state was chosen as an example of the valence band state due to its large photoemission yield. The data for the HS state were recorded at room temperature, and those for the LS state were recorded at T=$140\,$K, i.e., well below the SCO transition temperature. For both spectroscopic features, the PEEM images of both spin states had to be recorded for two different electron energies to account for the characteristic energy shift during the HS-LS transition. The magnitude of the corresponding energy shift is illustrated in the corresponding total yield spectra of the valence band and the low energy cutoff region shown in Fig.~\ref{figure2}(e) and~(f). 

Overall, the PEEM images of the HOMO-2 emission show remarkable uniformity for both the HS and LS states, indicating a homogeneous distribution of molecules with the same spin state throughout the sample. The overall intensity difference in the PEEM images can be attributed to the different line shapes of the HOMO-2 state in both spin states as illustrated in the total emission spectrum in Fig.~\ref{figure2}(e). The PEEM images of the cutoff region show a combination of spatially homogeneous areas decorated with darker spots caused by defects in the substrate crystal. These structural defects strongly impact the sample's local work function or ionization threshold and thus show an enhanced contract in the low energy cutoff region.

To monitor the spatial and temporal dynamics of light-induced spin-state switching in this thin film, we start with a homogeneous sample in the LS state at $140\,$K and illuminate the sample with a continuous wave (cw) laser diode ($\lambda=532\,$nm) or fs laser pulses with almost identical photon energy ($\lambda=530\,$nm). For these experimental conditions, optical excitation leads to the trapping of the light-induced spin state, known as light-induced excited state strapping (or LIESST). This LIESST state can persist for several milliseconds to hours after laser excitation (depending on the precise sample temperature) and can be probed with a continuous wave VUV light source (He I$_\alpha$ radiation, h$\nu=21.2\,$eV).

\begin{figure}[t!]
 \centering
 \includegraphics[width=16.0cm]{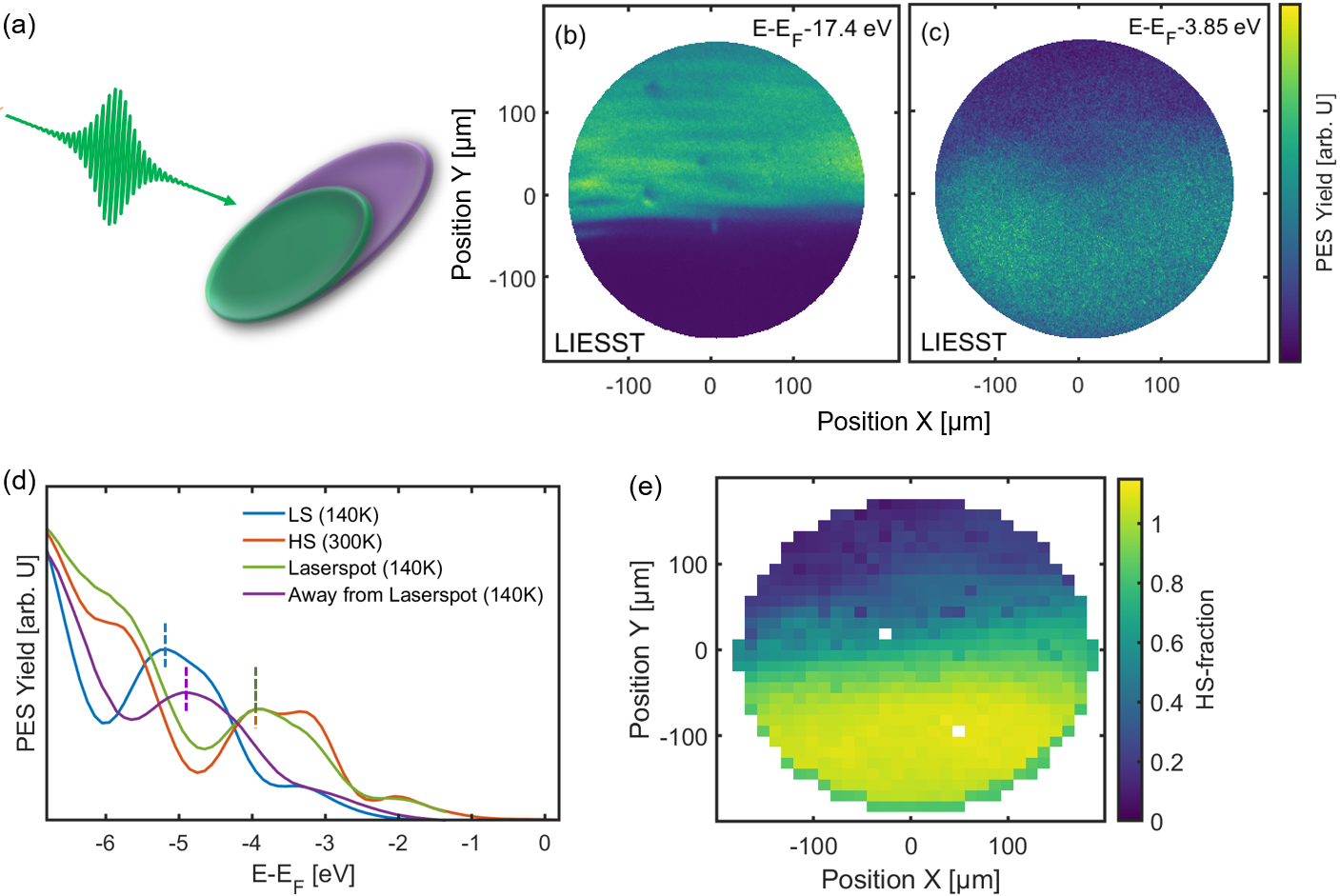}
 \caption{(a) Schematic representation of the experimental procedure. The laser spot (green area) of the cw laser diode is placed in the lower part of the field of view (FOV) of our PEEM instrument. PEEM images of the metastable HS-like LIESST state of a $40\,$ML Fe(phen)$_2$(SCN)$_2$ film on gold after excitation with a laser diode ($\lambda=532\,$nm) at $140\,$K. The images in (b) and (c) were taken for the low energy cutoff and the HOMO-2 energy region, respectively. (d) Spatially integrated valence band structure for the LS (blue solid line) and HS (orange solid line). In addition, we included the spatially resolved photoemission yield extracted in the directly laser illuminated (green solid line) and unilluminated (purple solid line) regions of the FOV. (e) Spatially resolved fraction of molecules in the LIESST state during laser irradiation. Note that a fraction of $\gamma_\mathrm{LIESST}=1$ indicates a complete conversion of SCO molecules to the LIESST state, while $\gamma_\mathrm{LIESST}=0$ indicates that all SCO molecules remain in the LS state.} 
 \label{figure3}
\end{figure}

Our experimental procedure is illustrated in Fig.~\ref{figure3}(a). The beam of a cw laser diode with wavelength $\lambda=532\,$nm is focused to a spot size of $\approx 100\,\mu$m on the sample surface and centered in the lower half of the field of view (FOV) of our PEEM instrument, as shown by the green area. In this way, only a certain part of the SCO film in the LS state within our FOV is directly optically excited by the laser beam into the LIESST state. The remaining part of the SCO film in the FOV (purple area) is not directly affected by the excitation and can only be switched to the LIESST state by intermolecular coupling and cooperative effects within the SCO film. By carefully analyzing the laser-illuminated area in the FOV, we gain insight into the cooperative nature of the spin-crossover transition, providing a nuanced understanding of the dynamic interplay within the molecular film.

Exemplary data of the spatial distribution of the light-induced switching behavior are shown in Fig.~\ref{figure3}(b) and (c) for two characteristic spectral features (low energy cutoff region and HOMO-2 feature). In both images, we observe an opposite bright/dark contrast in the optically excited and unexcited part of the FOV. This change in photoemission contrast can be directly correlated to the energy shifts of the spectral features during the HS/LS transition discussed above.
The PEEM image in Fig.~\ref{figure3}(b) was recorded in the low energy cutoff region between the maxima of the HS and LS spectra (see Fig.~\ref{figure2}(f)), which shows medium photoemission yield for the LS state but vanishing intensity for the HS state. Accordingly, the vanishing photoemission intensity indicates a direct transition from the LS to the HS-like LIESST state in the photoexcited region, while the SCO complexes appear to remain in the LS state in the region outside the laser spot.

Similar arguments can also qualitatively explain the contrast changes of the PEEM image in Fig.~\ref{figure3}(c) recorded for the binding energy of the HOMO-2 feature in the HS state. The light-induced LS to HS transition (LIESST state) shifts the maximum intensity of the HOMO-2 emission in the illuminated part of the FOV into the experimentally monitored energy window, while it remains unchanged in the optically unexcited region of the FOV. 

A more quantitative view of the spatial dependence of the light-induced spin state switching can be obtained by analyzing the spatially-resolved line shape of the valence band spectra within the entire FOV. Fig.~\ref{figure3}(d) shows the local valence band structure of the optically excited region (green curve, excited with cw laser diode) and the unexcited region (purple curve). 
The spectrum of the cw laser-illuminated region (green curve) almost perfectly matches that of the HS state recorded at room temperature (orange curve). Slight differences in the line shape of the HOMO-1/-2 emission (E-E$_\mathrm{F}\approx -3\,$eV) are again due to permanent VUV radiation-induced modifications of the SCO film. This quantitatively confirms the light-induced spin transition from the LS to the HS-like LIESST state even under illumination with a cw laser diode. Interestingly, we find significant differences between the local spectrum of the unilluminated region (purple curve) in Fig.~\ref{figure3}(d). In particular, the binding energy positions of the molecular features in this spectrum are between their positions for the LS and HS states without laser illumination. These observations suggest a partial spin-state switching of the SCO film in the vicinity of the laser illuminated area and thus hint to a cooperative expansion of the light-induced spin-state switching within the SCO film.

To further characterize the SCO transition in the unilluminated region, we quantify the light-induced shift of the HOMO and HOMO-2 signals using a dedicated fitting procedure. First, we extract local valence band spectra from our 3D dataset ($I(x,y,\mathrm{E-E_F})$) averaged over a small area consisting of $4\times 4$ pixels on the detector. This reduces the overall number of pixels in our images, but substantially improves the signal-to-noise ratio. We then fit the spectra in the energy range from HOMO to HOMO-4 using a Gaussian function per molecular state. To increase the reliability of our fitting results, we fix the relative energy positions of all Gaussian peaks, resulting in only a single fitting parameter for the global position of the valence band features. In our case, this is the binding energy of the HOMO-2 feature (E$_\mathrm{orbital}$). This relative shift of the valence band spectrum with respect to its position in the HS (E$_\mathrm{HS}$) and LS (E$_\mathrm{LS}$) states allows us to estimate the fraction between the molecules in the HS-like LIESST state and the LS state using the following equation: 
\[
\gamma_\mathrm{LIESST} = \frac{{E_\mathrm{\text{orbital}} - E_\mathrm{LS}}}{{E_\mathrm{HS} - E_\mathrm{LS}}}
\]
A ratio of $\gamma_\mathrm{LIESST}=1$ indicates that all SCO complexes are in the LIESST state, while $\gamma_\mathrm{LIESST}=0$ indicates that all molecules are in the LS state. 

For the HOMO-2 feature we determined E$_\mathrm{LS}=5.30\,$eV and E$_\mathrm{HS}=-3.98\,$eV for the unilluminated sample. Together with the fitting result E$_\mathrm{orbital}$, we calculate the relative ratio $\gamma_\mathrm{ LIESST }$ for the laser-illuminated sample and plot its spatial distribution in Fig.~\ref{figure3}(e). The cw laser-illuminated region in the lower part of the FOV yields a $\gamma_\mathrm{LIESST}$ of $\approx 1$, indicating a homogeneous spin transition from the LS to the LIESST state for all SCO complexes. Near the laser-excited region, $\gamma_\mathrm{LIESST}$ decreases but remains greater than zero even at distances of about $200\,\mu$m from the edge of the cw laser-illuminated region. Even at these large distances, we find $\approx 20\,\%$ of the SCO molecules in the LIESST state. This observation highlights the role of cooperative interactions between SCO molecules in the spatial distribution of light-induced spin state switching. We attribute this observation to 
the spatial temperature gradients in the SCO film after cw laser excitation. These gradients lead to heat transport from the optically excited to the unexcited region, causing thermally driven spin switching in the optically unexcited regions. Remarkably, this leads to cooperative effects on the SCO transition that extend up to distances of 200~$\mu$m and decrease with distance from the initially switched molecules.

\begin{figure}[ht]
 \centering
 \includegraphics[height=9.5cm]{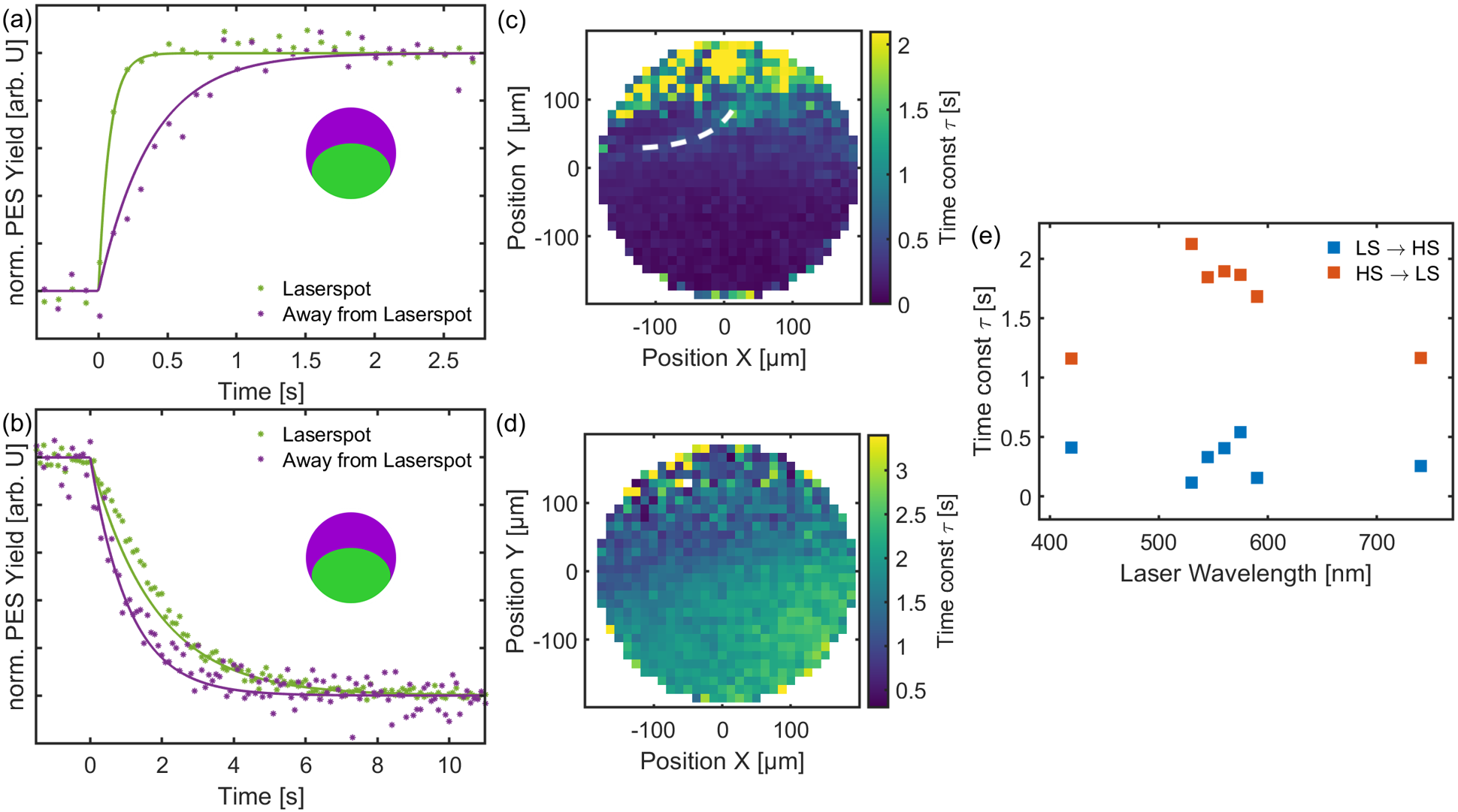}
 \caption{(a) Temporal evolution of the photoemission intensity of the HOMO-2 feature (E-E$_\mathrm{F}=-3.85\,$eV) of a $40\,$ML Fe(phen)$_2$(SCN)$_2$ film ($T=130\,$K) after laser excitation at $532\,$nm in selected regions of the FOV. The data points were obtained by integrating the photoemission yield in the laser-illuminated (green data points) and unilluminated (purple data points) regions of the FOV shown as an inset. The solid line represents an exponential fit to the data to determine the characteristic timescale of the LS to LIESST transition. Panel (b) shows similar spatially resolved photoemission data of the LIESST to LS state relaxation after blocking the laser excitation with a mechanical shutter. The spatially resolved time constants of the LS to LIESST and LIESST to LS transitions are shown in panels (c) and (d). (e) shows the time constants for the LS to LIESST (blue data points) and LIESST to LS (orange data points) transitions after excitation with fs laser pulses of different photon energies.}
 \label{figure4}
\end{figure}

More insight into the cooperative SCO switching in this thin $40\,$ML Fe(phen)$_2$(SCN)$_2$ film can be obtained by accessing the entire spatial and temporal evolution of the light-induced spin state switching from the LS to the HS state. Due to the light-induced formation of the LIESST state, we cannot access the initial femtosecond dynamics of the spin state switching in this material by pump-probe spectroscopy \cite{femtosecondXANES, ultrafastsco,Kammerer+2024}, which would require a full relaxation of the excited spin state within a few $\mu$s. Instead, we focus on the spatial and temporal dynamics of the spin-state switching associated with local thermal heating by cw laser illumination. In particular, we want to monitor the temporal evolution of the spin state switching in the unilluminated regions of the sample discussed above. 

To this end, we record a series of PEEM images of the at Fe(phen)$_2$(SCN)$_2$ film in the LS state ($T=130\,$K) at the binding energy of the HOMO-2 feature (E-E$_\mathrm{F}=-3.85\,$eV) at a frame rate of $10\,$frame/s or one image every $100\,$ms. The cw laser was again directed to illuminate only the lower part of the FOV, as shown in Fig.~\ref{figure3}(a). The illumination of the sample is controlled by a manual shutter with an opening and closing time of $<0.1\,$s. 
Fig.~\ref{figure4}(a) shows the temporal evolution of the intensity changes in the laser-illuminated and non-illuminated areas within the selected field of view in green and purple, respectively. Illumination starts at $t=0\,$s by opening the laser shutter. We observe an instantaneous increase in intensity in the laser-illuminated area. As discussed above, the intensity increase is caused by the valence band shifts during the transition from LS to LIESST. The optical response must be instantaneous with respect to our temporal resolution since the direct light-induced SCO transition is expected to occur on femto- to picosecond timescales \cite{femtosecondXANES, ultrafastsco,Kammerer+2024}. Using an exponential fit function (solid line), we determine a rise or switching time of $0.08\,$s, which we take as an estimate for the temporal resolution of our experimental procedure. 

The transition time of the non-illuminated regions is of much greater interest. We find an average switching time of $\tau=0.38\pm 0.08\,$s which is slower than in the directly illuminated area. This response time is in the same order of magnitude as those of purely thermally induced switching dynamics in SCO single crystals, where time constants as fast as $\tau=0.15\,$s have been reported \cite{spatiotemporalcrystal}. 

The complete spatial and temporal dynamics can be obtained by quantifying the switching times $\tau$ over the entire FOV. Therefore, we use a binning of $4\times 4$ pixel in our PEEM images to extract and fit each pixel's rise time. The spatial distribution of the switching times $\tau(x,y)$ is shown in Fig.~\ref{figure4}(c). In this plot, we find the fastest switching times ($\tau(x,y)\approx 0.1\,$s) in the laser-illuminated area in the lower part of the FOV. The switching times increase with increasing distance from the excitation area and reach values of up to $\tau(x,y)=1.50\pm 0.08\,$s. Even larger values of $\tau(x,y)\approx 2\,$s are observed at the upper edge of the FOV. These values are, however, less reliable due to the low signal-to-noise ratio of the spectra in this region. 

Besides the general increase of the switching time with distance from the excitation area, we also find switching time hotspots in the unilluminated area. An example is the ark-like feature in the upper half of the FOV, highlighted by the white dashed line. These hotspots or hotspot regions are attributed to local structural inhomogeneities in the disordered SCO film. 
Despite these inhomogeneities of our SCO film, we estimate the average speed of the phase front in the LS/LIESST state after the light-induced spin-state switching. To do this, we use the maximum propagation range of the phase front and divide it by the average switching time constant for that range. The resulting velocity of the phase front is v$_\mathrm{PF}=526\pm 30\,\mu$m~s$^{-1}$. Remarkably, this average speed in the disordered SCO thin film is comparable to the reported values for the spin transition in [Fe(HB(tz)$_3$)$_2$] single crystals \cite{spatiotemporalcrystal} and exceeds the reported values of most other single crystal studies \cite{photothermal, propagationthermal}.

Besides the light-induced SCO transition, we also focused on the relaxation time of the metastable LIESST state back to the LS state. For a quantitative analysis, we recorded a series of PEEM images after closing the laser shutter using identical parameters as described above. Fig.~\ref{figure4} (b) shows spatially integrated time traces for the illuminated (green) and unilluminated (purple) parts of the FoV. We find small differences in relaxation between the two regions, with a slightly faster decay for the optically unexcited region of $\tau_R=1.08\pm 0.08\,$s (vs. $\tau_R=1.79\pm 0.08\,$s in the directly illuminated region). 
The full spatial distribution of relaxation times is shown in Fig.~\ref{figure4}(d) and reveals the same spatial inhomogeneities as discussed above for light-induced switching. 

Overall, the relaxation of the LIESST state into the LS state is about 10 times slower than the laser-induced LS to LIESST transition. Similar ratios have been observed for purely thermal switching of SCO single crystals. Depending on the material system, the ratio of LS to HS and HS to LS transition times varies from 1:10 \cite{photothermal} to 1:2 \cite{spatiotemporalcrystal}. This variation appears to be influenced by factors such as crystal size, crystal quality, and proximity of the temperature to the transition temperature.

The quantitative differences in the switching and relaxation time constants for different excitation wavelengths are shown in Figure \ref{figure4}(e). The different wavelengths of the laser radiation were generated by an optical parametrical oscillator, which emits ultrashort light pulses with pulse duration of $\approx 200\,$fs and average power of $6\,$mW (at $80\,$MHz repetition rate). 
The time constants are again determined by an exponential fit to the local photoemission yield in the directly photoexcited part of the FOV. The blue data points represent the SCO switching time constant from the LS to the LIESST state. The switching times scatter between $\tau=0.1\,$s and $\tau=0.5\,$s and do not show a clear maximum in the investigated wavelength range. This is consistent with a purely light thermally driven spin switching process. On average, the switching times observed for fs laser excitation are longer than those determined for cw laser excitation. We attribute this difference to a reduction in the signal-to-noise ratio in the fs laser excitation data and, hence to a greater uncertainty in the experimental values compared to the laser diode experiments. 
In contrast, the relaxation times show values between $\tau_R=1\,$s and $2.1\,$s with a clear maximum at $530\,$nm. We attribute the maximum in the relaxation times around $530\,$nm to the energy deposited in the SCO film and the corresponding light-induced temperature rise in the SCO film. The wavelength with the largest relaxation times coincides with a maximum of light absorption of such films in the range between $500\,$nm and $600\,$nm (see \citet{uhvsco1}). Accordingly, the increased energy input for these wavelengths leads to a more significant energy content in the molecular film for these wavelengths on longer time scales, which is still sufficient to stabilize the LIESST state thermally. Correspondingly, the LIESST state can only return to the LS state if this enhanced energy content has dissipated.

\section*{Conclusion}
In conclusion, our study has elucidated the spatial and temporal dynamics of the collective spin state switching of nanometer thin SCO films on a gold single crystal. For the exemplary case of the Fe(phen)$_2$(SCN)$_2$ compound, we have followed the spatial and temporal evolution from the low spin (LS) to the high spin (HS)-like LIESST state using real-time photoemission electron microscopy with millisecond time resolution. Our comprehensive characterization of the spectroscopic photoemission signatures of the temperature-dependent LS and HS states, as well as their changes during the temperature- and light-induced spin transition, allowed us to quantify the LS to LIESST state switching ratio with (sub)-$\mu$m precision. 

We uncover spectroscopic signatures of the light-induced spin state switching to the LIESST state even $200\,\mu$m away from the spot of direct optical excitation. This observation is identical for both continuous wave and femtosecond laser light sources and can be observed in the spectral range between $\lambda=420$ - $740\,$nm. We attribute this cooperative interaction between the SCO molecules largely to a photothermal process caused by the heat transport from the optically excited to the unexcited region leading to thermally driven spin switching in the optically unexcited regions.

Overall, our study has demonstrated the critical role of intermolecular interactions, cooperativity, and thermal effects in local light-induced spin switching in thin SCO films on surfaces. These phenomena set limits to the local confinement of optically controllable spin switching, with potential implications for light-driven nano- or micrometer-scale spintronic device structures based on functional SCO materials. 

\section*{Experimental Methods}

All experiments were conducted in an ultrahigh vacuum (UHV) system consisting of three separate chambers: an analysis chamber, a surface preparation chamber, and an evaporation chamber. The base pressure in all chambers was consistently better than $3 \times 10^{-10}$ mbar.

The Au(111) surface was prepared by repeated cycles of argon ion sputtering and sample annealing at $T=800\,$K for 30~min.  Subsequently, Fe(phen)$_2$(SCN)$_2$ films with thicknesses ranging from 0.5 to 50 monolayers were deposited on the pristine Au(111) surface at room temperature by organic molecular beam epitaxy using a commercial Knudsen cell evaporator (Kentax GmbH) filled with Fe(phen)$_2$(SCN)$_2$ powder. The coverage of Fe(phen)$_2$(SCN)$_2$ films was estimated by monitoring deposition time and rate using a quartz microbalance and calibrated by the intensity of the C-KLL and Au-NOO Auger peaks. After deposition, Fe(phen)$_2$(SCN)$_2$ films were annealed at 420~K to produce high quality films.  This ensures the reversion of Fe(phen)$_2$(SCN)$_2$ to Fe(phen)$_2$(SCN)$_3$, a reaction that occurs during evaporation due to the elevated temperatures \cite{uhvsco1}. 

Spatially averaged photoemission spectroscopy (ARPES) experiments were performed using a hemispherical analyzer (SPECS Phoibos 150) with a CCD detector system for two-dimensional detection of electron energy and momentum. Monochromatic He I$\alpha$ radiation (21.2~eV) from a high-flux He discharge source (VUV 5000) was used. Due to lateral disorder of the molecules, all spectra were angle integrated over a $\pm$13° ($\pm$0.48 Å$^{-1}$) acceptance angle around the $\Gamma$ point of Au(111). The pass energy was set to 10~eV and the angular resolution was 0.3°. Samples were cooled to 15~K using the integrated helium cryostat.

Photoemission electron microscopy (PEEM) images were obtained using a double hemispherical energy filter photoemission electron microscope (NanoEsca system). We used an extractor voltage of 200~V, resulting in a magnification of 130 and a field of view of $\approx 300\,\mu$m. The sample was cooled with liquid nitrogen to 140~K, and illumination was provided by a He discharge source (Specs UVS 300).

The light-excited spin-crossover transition was induced using a diode laser (532~nm) with an intensity of 2.8~mW/mm$^2$, illuminating the sample simultaneously with the photoemission experiments. Additionally, femtosecond pulses with pulse durations of $\approx 200\,$fs were generated using an optical parametric oscillator (Inspire from Spectra Physics) with wavelengths ranging from 740 to 420~nm. The intensity was varied between 9 and 900~mW/mm$^2$ (0.07-7~mW).

\begin{acknowledgement}
This work was funded by the Deutsche Forschungsgemeinschaft (DFG, German Research Foundation) - TRR 173 - 268565370 Spin + X: spin in its collective environment (Project A02 and A04). 
\end{acknowledgement}

\begin{suppinfo}

Experimental details for thickness dependent energy shifts and EDCs for the HS, LS and LIESST states of the molecular film near E$_\mathrm{Fermi}$.

\end{suppinfo}

\bibliography{achemso-demo}
\bibstyle{achemso}

\end{document}


\flushbottom
\maketitle
%
%

\renewcommand\thepage{S-1}

\section*{Thickness Dependent Binding Energy Position of the HOMO Peak}

 \renewcommand{\thefigure}{S1}
 \begin{figure}[ht]
\centering
\includegraphics[width=0.7\linewidth]{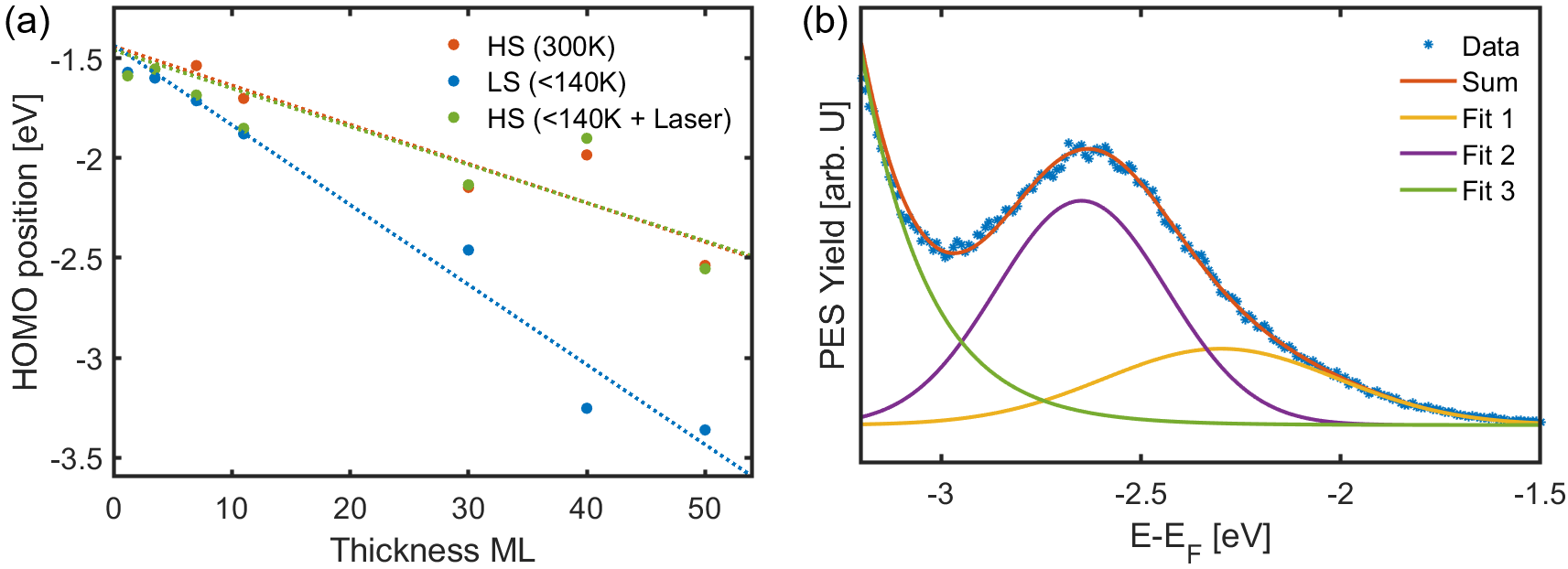}
\caption{(a) Fitting result (binding energy position) of the HOMO peak of the thin Fe(phen)$_2$(SCN)$_2$ films on Au(111) for different film thicknesses and in different spin states, i.e. in the LS, HS and LIESST states. The binding energy positions were determined by a dedicated lineshape analysis of the valence band photoemission data using the fitting model shown in panel (b).}
\label{fig:SI2}
\end{figure}

Figure \ref{fig:SI2}~(a) shows the fitted binding energy positions of the highest occupied molecular orbital (HOMO) feature for all film thicknesses obtained for all accessible spin state: the high spin (HS), the low spin (LS) and light induced excited spin state trapping (LIESST) state. The corresponding fitting model is shown in Fig.~\ref{fig:SI2}(b). The valence band spectra are fitted with two Gaussian functions (labeled Fit 1 and Fit 2) and an additional exponential function (labeled Fit 3) to model the onset of the HOMO-1 peak.
In the HS state at room temperature, the binding energy position of the HOMO (i.e. the absolute value of the HOMO position) increases continuously with increasing coverage. The binding energy of the LS state is always greater than that of the HS state. The energy difference between the HS and LS states (recorded at T$=140\,$K) increases monotonically with increasing film thickness, from only $18\,$meV at $1.2\,$ML to $820\,$meV at $50\,$ML. Crucially, the light induced spin state change from the LS state to the LIESST state at $140\,$K leads to a shift of the HOMO feature from its position in the LS state to its position in the HS state at room temperature. This shows that the binding energy position of the HOMO (and the subsequent HOMO-X features) can be considered as a fingerprint of the thermal and optical spin state switching of the SCO complexes.

\clearpage
\renewcommand\thepage{S-2}
\section*{Spectral lineshape of the valence bands of thin films of the Fe(phen)$_2$(SCN)$_2$ complex for the HS, LS, and LIESST states}

\renewcommand{\thefigure}{S2}
\begin{figure}[ht]
\centering
\includegraphics[width=0.7\linewidth]{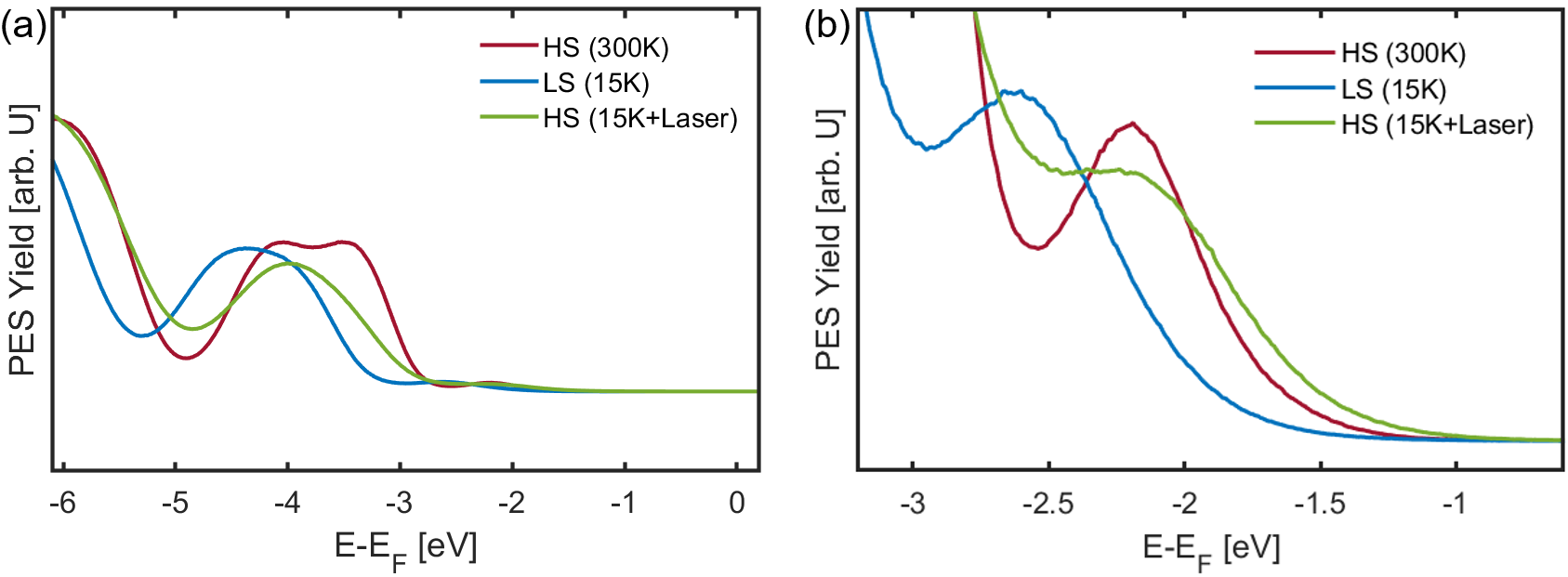}
\caption{(a) Photoemission valence band spectra (E$_{photon}=21.23\,$eV) of a $30\,$ML film of Fe(phen)$_2$(SCN)$_2$ on Au(111) recorded for the HS (room temperature), LS ($15\,$K) and LIESST state ($15\,$K, excited with cw laser source). A close up of the HOMO region of all three spectra is shown in panel (b).}
\label{fig:SI1}
\end{figure}

Figure~\ref{fig:SI1}(a) shows the overview valence band spectra of a $30\,$ML film of Fe(phen)$_2$(SCN)$_2$ on an Au(111) single crystal in the HS state at room temperature (red solid line) and in the LS state at $15\,$K. In addition, the valence band spectra of the LIESST state at $15\,$K are shown as a green solid line. The LIESST state was excited using a cs laser diode ($\lambda=532\,$nm) with a relatively low laser fluence of $2.8$mW$\,$mm$^{-2}$. A close-up of the valence band spectra in the region of the HOMO peak is shown in Figure~\ref{fig:SI1}(b).
A direct comparison of these spectra shows an energy shift of $\approx 0.3\,$eV towards higher binding energies when the sample is cooled from room temperature to $15\,$K. Upon laser excitation, this energy shift is almost completely reversed, indicating the light-induced transition from the LS to the HS-like LIESST state. The formation of the LIESST state is observed for both continuous wave and pulsed laser light sources. The spectral broadening of the HOMO level is attributed to radiation-induced degradation of the SCO film due to the exposure to VUV light required for the photoemission process.

